\documentclass[aps,pra,preprint,superscriptaddress, groupedaddress]{revtex4-1}

\usepackage{graphicx}
\usepackage{dcolumn}
\usepackage{bm}
\usepackage{amssymb}
\usepackage{amsmath}
\usepackage{amsthm}
\usepackage{color}

\usepackage{hyperref}
\hypersetup{
	colorlinks=true,
	linkcolor=blue,
	filecolor=magenta,      
	urlcolor=cyan,
	pdftitle={Overleaf Example},
	pdfpagemode=FullScreen,
}

\newtheorem{theorem}{Theorem}
\newtheorem{lemma}{Lemma}

\begin{document}


\title{Optimal local filtering operation for enhancing quantum entanglement}

\author{Zhaofeng Su}
\altaffiliation{These authors contributed equally to this work.}
\affiliation{School of Computer Science and Technology, University of Science and Technology of China, Hefei 230027, China.} 

\author{Nina Sukhodoeva}
\altaffiliation{These authors contributed equally to this work.}
\affiliation{School of Computer Science and Technology, University of Science and Technology of China, Hefei 230027, China.}

%


\date{\today}

\begin{abstract}
  Quantum entanglement is an indispensable resource for many significant quantum information processing tasks. Thus, distilling more entanglement from less entangled resource is a task of practical significance and has been investigated for decades. The literature [Verstraete \textit{et al}., \href{https://link.aps.org/doi/10.1103/PhysRevA.64.010101}{Phys. Rev. A 64, 010101(2001)}]  considered a scenario to increase the entanglement by local filtering operation and qualitatively  derived the variance relation of entanglement. We investigate the scenario with general two-qubit resources to find the optimal strategy of filtering operations. We  obtain the upper bound for the ratio of entanglement increase and find the corresponding optimal local filtering operation to achieve the maximal ratio. Our analysis shows that the upper bound ratio grows with the length of local Bloch vector while the success probability decrease with it. We further extend the research to investigate the optimal measurement strategy by considering general measurement. Our result shows that local measurement can not increase the expectation of quantum entanglement, which gives more analytical evidence to the well known fact that local operation can not create quantum entanglement. 
  
\end{abstract}

\maketitle

\section{Introduction}
\label{intro}
The phenomenon of the existence of non-classical correlations between spatially separated quantum systems is known as entanglement, which is firstly noted by well known EPR paradox~\cite{Einstein1935} and named by Schr{\"o}dinger~\cite{Schrodinger1935} . Together with nonlocality~\cite{Bell64} , entanglement is the fundamental property of quantum mechanics. Those remarkable properties of quantum mechanics have fundamentally changed human’s view of the world and taught us to think the unthinkable. The combination of quantum mechanical theory and computation theory provides a new paradiagram for computing, which is known as quantum computation and has exponential advantages over its classical counterpart~\cite{Nielsen2000}. Quantum entanglement is acted as the indispensable resource in many important protocols of quantum computation and information tasks~\cite{RH09}. For example, the important protocols quantum dense coding and quantum teleportation are based on the resource of maximally entangled states~\cite{PhysRevLett.69.2881, PhysRevLett.70.1895} . Protocols are also proposed to generate maximal entanglement from general entangled states~\cite{ZS17} and to extend transmission distance via efficient strategy of quantum repeaters with respect to the transmission rate and the cost of local operation and classical information (LOCC)  ~\cite{PhysRevA.97.012325}. Therefore, quantum entanglement do not only has theoretical importance on quantum mechanics but also has practical significance in quantum computation and information.

It is a well-known fact that separable quantum states do not contain any entanglement and can not be used for quantum advantage protocols. On the contrary, singlet states have widely used in advantage quantum protocols and known as maximally entanglement. And there are other states containing some amount of quantum entanglement between separable states and singlet states. Therefore, it is of practical significance to measure the quantity of entanglement for a specified quantum system. Several measures of quantum entanglement have been proposed in last three decades. For example, distillable entanglement~\cite{Plenio2007}, entanglement of formation (EoF)~\cite{PhysRevA.54.3824} and concurrence~\cite{PhysRevLett.78.5022}.   For the two-qubit system,  EoF and concurrence are acted as the similar measure of entanglement but with different scale, which are related by the binary entropy function.  

In this work, we consider the concurrence as the measure for quantum entanglement of a general two-qubit system.  Analytically, the concurrence $C(\rho)$ of a general two-qubit state $\rho$ is equal to the following equation,
\begin{equation}
	C(\rho) = max\{ 0, \lambda_1 - \lambda_2 - \lambda_3 - \lambda_4 \},
\end{equation}
where $\lambda_i$ are eigenvalues of the operator $R\equiv
\sqrt{\sqrt{\rho}\Tilde{\rho}\sqrt{\rho}}$ in decreasing order and $\Tilde{\rho}=(\sigma_y\otimes \sigma_y)\rho^*(\sigma_y\otimes \sigma_y)$. Then, the EoF of $\rho$ can be calculated via the relation $EoF(\rho) = h \left( \frac{ 1 + \sqrt{1-C(\rho)^2}} {2} \right)$ where $h(x) \equiv -x\log_2 x - (1-x)\log_2 (1-x)$  is known as Shannon's binary entropy function~\cite{PhysRevLett.80.2245}. 

Due to the necessity of maximally entangled resource in the quantum information protocols, it is of practical significance to create maximally entangled states from less entangled resources, which is knowns as quantum entanglement distillation~\cite{PhysRevA.67.022310}.  It is well known that quantum entanglement is a resource that  could not be created via local operation and classical communication (LOCC). However, it is possible to enhance quantum entanglement by local operations, say generating maximally entangled state from less entangled resources by local operations. One of such local operations is filtering operation. Suppose Alice and Bob share a two-qubit state $\rho_{AB}$. The filtering operation $F$ applies on any side, say Alice,  is an operator on the subsystem with restriction $F^{\dagger} F \le I$. The filtering operation evolutes the joint system as following,
\begin{equation}
	\rho_{AB}' = \frac{(F \otimes I)\rho_{AB} (F \otimes I)^{\dagger} }{tr((F^{\dagger}F \otimes I )\rho_{AB}) }.
\end{equation}
Local filtering operations are widely used in quantum information processing protocols. The filtering operation is useful for teleportation applications  and the optimal solution with respect to the fidelity were proposed~\cite{PhysRevLett.90.097901}. The distillable key rate for quantum key distribution can be increased by local filtering operations~\cite{PhysRevA.102.032415}.
The filtering operation has been successfully realized by physical experiments~\cite{PhysRevA.65.052319, ChaoLiu2022}.

Verstraete \textit{et. al.}~\cite{PhysRevA.64.010101} investigated the effect of local filtering operation on two-qubit state and found the following quantitative relation,
\begin{equation}
	C( \rho_{AB}' ) = C( \rho_{AB}) \frac{ |det(F)| }{tr((F^{\dagger}F \otimes I )\rho_{AB}) },
\end{equation}
which provides the dynamic change of entanglement over general local filtering operation.  However, it is still an open question to analytically find the optimal filtering operation which can maximally enhance the quantum entanglement with respect to  concurrence for a given quantum state. 

In this paper, we consider the effect of local operations on general two-qubit system and further investigate the open problem remained in the work of Verstraete \textit{et. al.} ~\cite{PhysRevA.64.010101}.  In Sec.~\ref{sec:optimalFO}, we try to find an analytical expression of the optimal local filtering operation with respect to the maximal entanglement increase measured by concurrence. In Sec.~\ref{Sec:effectofMeasurement}, we make extension to consider the effect of local general measurements for quantum entanglement and give numerically evidence to the well-known fact that local measurement can not increase the expectation of quantum entanglement.

\section{Optimal filtering operation}\label{sec:optimalFO}
In this section, we investigate the optimal filtering operation for generating entanglement from a general two-qubit system and further consider the effect of local measurement.

Consider the scenario that  separated parties Alice and Bob share a  two-qubit quantum system. Suppose $\rho_{AB}$ is the general state of the system,   which can be represented with respect to Pauli operators as follows,
\begin{equation}\label{eq:2qubitstate}
	\rho_{AB} = \frac{1}{4}(I \otimes I + \vec{a}\cdot \vec{\sigma} \otimes I + I \otimes \vec{b}\cdot \vec{\sigma} + \sum_{j_{1},j_{2}=1}^{3} T_{j_{1}j_{2}} \sigma_{j_1} \otimes  \sigma_{j_2}),
\end{equation}
where $\vec{a}, \vec{b}$ are vectors in $\mathbb{R}^3$ with elements $a_{j} = tr(\rho_{AB}(\sigma_j \otimes I))$ and $b_{k} = tr(\rho_{AB}(I \otimes \sigma_k))$ for $k,j\in\{1,2,3\}$, $\vec{\sigma}$ is vector of Pauli operators, and $T_{j_1j_2}  = tr(\rho_{AB}(\sigma_{j_1} \otimes \sigma_{j_2} ))$ are elements of a $3\times 3$ real matrix $T$. Note that $\rho_{A} \equiv tr_{B}(\rho_{AB}) = \frac{1}{2}(I + \vec{a}\cdot \vec{\sigma})$ is the state of Alice's partial system and $\vec{a}$ contains all the information for the system. And similarly $\rho_{B} \equiv tr_{A}(\rho_{AB}) $ is that of Bob's system. It is obvious that the norm of vectors  $\vec{a}$ and $\vec{b}$  are no greater than 1. The property $tr(\rho_{AB}^2) \le 1$ of a density operator naturally results the constraint $\| \vec{a} \|^2  + \| \vec{b} \|^2  + \| T \|^2 \le 3$ where $\| T \|^2 \equiv \sum_{j_{1},j_{2}=1}^3 T_{j_1j_2}^2$~\cite{entropy2306728}.

Suppose Alice applies a filtering operation $F$ on her system. The filtering operator $F$ has polar decomposition of the form $F=U_{F}K_{F}$ where $K_{F}\equiv \sqrt{F^{\dagger}F}$ and $U_F$ is a unitary based on the singular value decomposition of $F$. To be specific, suppose  $F = \sum_k f_k |\alpha_k\rangle \langle \beta_k|$  is the singular value decomposition of $F$,  then $U_F = \sum_k \frac{f_k}{|f_k|} |\alpha_k\rangle \langle \beta_k|$  where we set $\frac{f_k}{|f_k|} = 1$ for the cases with zero singular values. 

Let $F= \frac{1}{2}\sum_{k=0}^{3} x_{k}\sigma_{k} = \frac{1}{2}(x_{0} I + \vec{x} \cdot \vec{\sigma})$ with $\vec{x} =(x_1, x_2, x_3) \in\mathbb{R}^3$ and $x_{j} \equiv  tr(K_{F}\sigma_{j})$ for $j=0, 1,2,3$. It is trivial to calculate that the eigenvalues of $K_F$ are $\frac{1}{2}(x_{0} + \|\vec{x} \|)$ and $\frac{1}{2}(x_{0} - \|\vec{x} \|)$.  Note that $K_F$ is a  positive operator and the filtering operation is constrained by $F^{\dagger}F\le I$. The parameters have to satisfy the  constraints $0 \le \frac{1}{2}(x_{0} - \|\vec{x} \|) \le \frac{1}{2}(x_{0} + \|\vec{x} \|) \le 1$, which is equivalent to the following 
\begin{equation}\label{eq:xnorm}
	\| \vec{x} \| \le x_{0} \le 2 - \| \vec{x} \|,
\end{equation}
with $\| \vec{x} \| \le 1$.

According to Verstraete's result~\cite{PhysRevA.64.010101}, the problem of find optimal filtering operation is equivalent to maximizing the ratio $ \frac{ |det(F)| }{tr((F^{\dagger}F \otimes I )\rho_{AB}) }$ over all possible filtering operator $F$. The determinant of filtering operator $F$ is 
\begin{equation}
	|det(F)| = |det(U_{F}) \cdot det(K_{F})| = |det(K_{F})|  = \frac{1}{4}(x_{0}^2 - \| \vec{x} \|^2).
\end{equation}
Note that $F^{\dagger}F = K_{F}^{\dagger}K_{F} = K_{F}^2$. Via some trivial calculation, we get the probability of observing the filtering operation as follows,
\begin{equation}\label{eq:probabilityFO}
	p(F) = tr((F^{\dagger}F \otimes I )\rho_{AB}) = \frac{1}{4} (x_0^2 + \| \vec{x} \|^2 + 2 x_0 \vec{x} \cdot \vec{a}).
\end{equation}
Let $\vec{\omega} \equiv \frac{\vec{x}}{x_0}$ and $\omega \equiv \| \vec\omega \|$ denote the corresponding vector norm. The ratio can be equivalently written as
\begin{displaymath}
	\begin{array}{ccl}
		\frac{ |det(F)| }{tr((F^{\dagger}F \otimes I )\rho_{AB}) } & = & \frac{1 - \omega^2}{1 + \omega^2 +2 \vec{\omega} \cdot \vec{a}} \\
		& \le & \frac{1 - \omega^2}{1 + \omega^2 - 2v_{A}a},
	\end{array}
\end{displaymath}
where the equality holds when $\vec{\omega} = -\frac{\omega}{a}{\vec{a}}$ and we have taken the notation $a\equiv \| \vec{a} \|$. To optimize over the possible filtering parameters, we define the function $f(\omega) \equiv  \frac{1 - \omega^2}{1 + \omega^2 - 2a\omega}$. The derivatives of the function is $f(\omega)' = \frac{2(a\omega^2 - 2\omega + a)}{(1 + \omega^2 - 2a\omega )^2}$. It is trivial to find that the equation $f(\omega)' = 0$ holds at points $\omega _+ = \frac{1 +  \sqrt{1-a^2}}{a}$ and $\omega _- = \frac{1 -  \sqrt{1-a^2}}{a}$  . Note that the variable $\omega \le 1$ due to the constraint of Eq.~(\ref{eq:xnorm}). Thus, the function have a sole valid extreme point  $\omega_- = \frac{1 - \sqrt{1-a^2}}{a}$ with the corresponding maximum value $f(\omega_-) = \frac{1}{\sqrt{1 - a^2}}$. Therefore,  the ratio that  a  filtering operation can increase the concurrence of the joint system has the upper bound as follows,
\begin{align}
	\frac{ |det(F)| }{tr((F^{\dagger}F \otimes I )\rho_{AB}) } \le \frac{1}{\sqrt{1 - a^2}}.
\end{align}
Note that the degree of mixture of the  state $\rho_{A}$ is characterized by its purity $P(\rho_{A}) \equiv tr(\rho_{A}^2) = \frac{1}{2}(1 + a^2)$~\cite{JPAMT43.5.055302}. Obviously, the closer the state $\rho_{A}$  to be pure state, the more entanglement of the joint system can be promoted.   

Now, we consider the format of the optimal filtering operation that can derive the maximal concurrence. Recall that any filtering operation is described by four parameters $x_0\in \mathbb{R}$ and $\vec{x}\in \mathbb{R}^3$. According to the aforementioned discussion, the optimal filtering operation should satisfy the constraints $\omega = \frac{1 - \sqrt{1 - a^2}}{a}$ and $\vec{\omega}   =   -\frac{ \omega}{a} \vec{a}$ where we have taken the notation $\vec{\omega} \equiv \frac{\vec{x}}{x_0}$.   It is obvious to conclude that a filtering operation $F$ can derive the optimal concurrence if  the parameters satisfy the following relation
\begin{align}\label{eq:MaximalConcurrenceCondition}
	\vec{x} = -x_{0} \frac{1 - \sqrt{1 - a^2}}{a^2} \vec{a}.
\end{align}
And the probability to observe the optimal operation is given by
\begin{align}
	p(F)  = x_0^2 \frac{(1-a^2)(1 - \sqrt{1-a^2})}{2a^2}.
\end{align}
The aforementioned discussion shows that the filtering operation can derive the maximum concurrence if the vector $\vec{x}$ is along  vector $\vec{a}$ of the local state while the scaling parameter $x_0$ affect the probability of observing the operation. To construct an optimal filtering strategy, we need to maximize the scaling parameter $x_0$ while the constraint in Eq.~(\ref{eq:MaximalConcurrenceCondition}) holds. Note that a  filtering operation $F$  is valid if the condition in Eq.~(\ref{eq:xnorm}) holds. Namely, $x_{0}\ge  \| \vec{x} \| \ge 0$ and $x_{0} + \| \vec{x} \| \le 2$. Further we can derive the constraint of $x_0$ as follows,
\begin{align}
	x_{0} \le 1 + \sqrt{\frac{1 -a}{1+a}}.
\end{align}
Therefore, the parameters for describing the optimal filtering  apparatus are $x_{0} = 1 + \sqrt{\frac{1 -a}{1+a}}$ and $\vec{x} = -(1 - \sqrt{\frac{1 -a}{1+a}})\frac{\vec{a}}{a} $. And the corresponding probability is $ 1 -a$. In the above discussion, we analyze the filtering operation of the polar decomposition form $F = U_{F}K_{F}$. Since local unitary operation has no effect on quantum entanglement, we only need to consider the positive operation $K_{F}$.

\begin{figure}[h] 
	\includegraphics[scale=0.4]{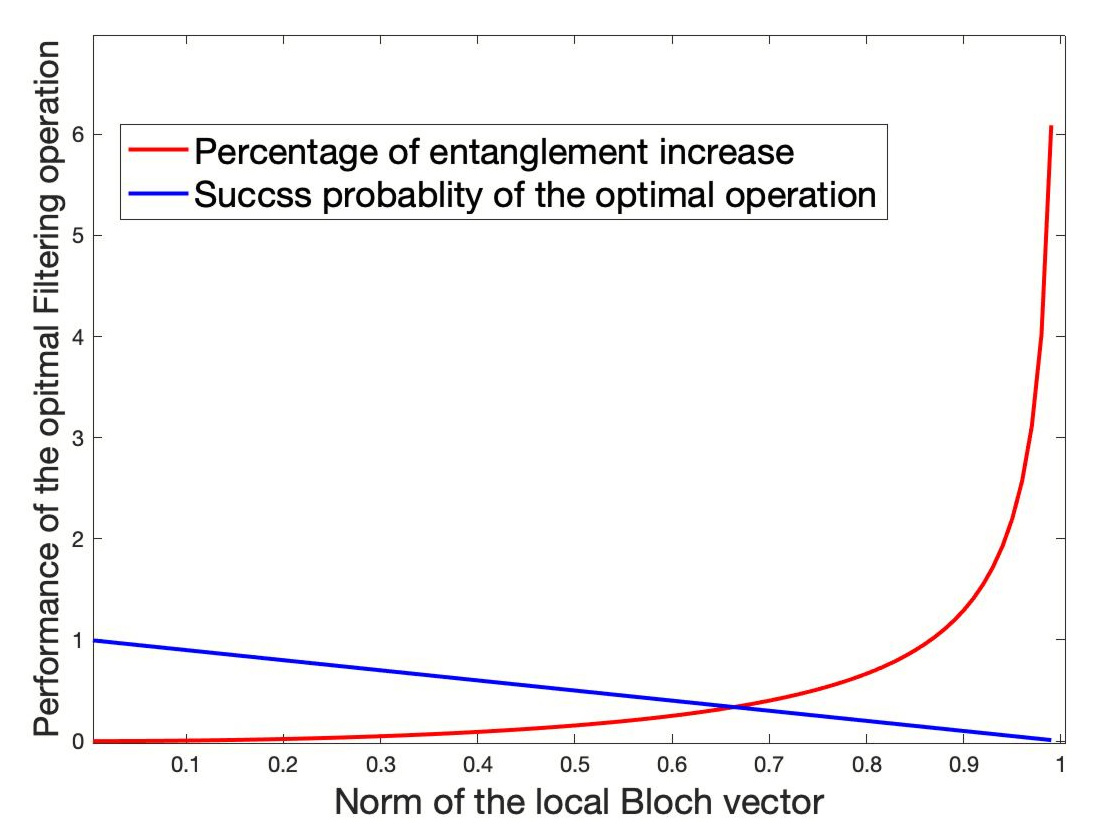}   
	\caption{The relationship between the local Bloch vector and the performance of the optimal Filtering operation.  }\label{fig:performance}
\end{figure}

It is obvious that the local Bloch vector of the qubit state plays a key role for determining the performance of the optimal Filtering operation.  The optimal Filtering operation can lead to entanglement increase with a ratio which grows with the norm of the local Bloch vector. On the contrary, the success probability of the optimal Filtering operation decrease with the Bloch vector norm. Note that the percentage of the entanglement increase is $\frac{1}{\sqrt{1 - a^2}} - 1$. We depict the performance in Fig.~(\ref{fig:performance}). 

We have proved the following theorem.
\begin{theorem}
	Suppose the general two-qubit state $\rho_{AB}$ represented in Eq.~(\ref{eq:2qubitstate}) is shared by separated parties Alice and Bob. By applying local filtering operation on one side, say Alice, the concurrence of the joint system can be  increased by ratio up to  $\frac{1}{\sqrt{1 - a^2}}$. The optimal filtering operation to derive the maximal concurrence is $F_{optimal} = \frac{1}{2}(( 1 + \sqrt{\frac{1 -a}{1+a}}) I -(1 - \sqrt{\frac{1 -a}{1+a}})\frac{\vec{a}}{a}\cdot \vec{\sigma})$ and the probability to observe the operation is $p(F_{optimal}) = 1 -a$.
\end{theorem}

Note that the density operator $\rho_{A} = \frac{1}{2}(I + \vec{a} \cdot \vec{a})$ of the partial system have eigenvalues $\frac{1}{2}(1 \pm a)$ with corresponding projectors $\frac{1}{2}(I \pm \frac{\vec{a}}{a} \cdot \vec{\sigma})$. It is trivial to get $\rho_{A}^{-\frac{1}{2}} = \frac{\sqrt{1-a} + \sqrt{1+a}}{\sqrt{2}\sqrt{1-a^2}} (I - \frac{1 - \sqrt{1-a^2}}{a} \frac{\vec{a}}{a} \cdot \vec{\sigma})$ which is proportional to $F_{optimal}$. Namely,  $F_{optimal} =\sqrt{ \frac{1-a}{2}}  \rho_{A}^{-\frac{1}{2}}$. In some literatures, the operator $\rho_{A}^{-\frac{1}{2}}$ is used as a quantum operator. However, the eigenvalue of  $\rho_{A}^{-\frac{1}{2}}$ is greater that 1 for mixed state $\rho_{A}$ and thus it can not act as a quantum operation. To be a valid quantum operator, a coefficient should be added such that $\mu \rho_{A}^{-\frac{1}{2}}$  with $\mu \le \sqrt{ \frac{1-a}{2}}$.

Further we consider the scenario that both Alice and Bob take local filtering operations. Suppose Bob's filtering operator is described by the operator $E = \frac{1}{2} \sum_{j=0}^3 y_{j} \cdot \sigma_{j}  =  \frac{1}{2} (y_0 I + \vec{y} \cdot \vec{\sigma})$. The filtering operations on both side can increase the entanglement measured by concurrence with ratio $\frac{|det(F)|\cdot |det(E)|}{tr((F^{\dagger} F \otimes E^{\dagger} E) \rho_{AB})}$. It is not obvious to optimize the ratio over all valid filtering operators $F$ and $E$ simultaneously. However, we can consider that there is a tiny gap in time to apply the operations. Alice applies the optimal filtering operation $F_{optimal}$ on her system and increase the concurrence by ratio $\frac{1}{\sqrt{1-a^2}}$, which does not affect the Bloch vector $\vec{b}$ of Bob's system. Nearly simultaneously, Bob applies the optimal filtering operation $E_{optimal}$ which is only determined by the Bloch vector $\vec{b}$ on his side and further increase the concurrence with ratio $\frac{1}{\sqrt{1-b^2}}$. We have taken the notation $b \equiv \| \vec{b} \|$. Eventually, the concurrence of the joint system is increased by ratio $\frac{1}{\sqrt{(1-a^2)(1-b^2)}}$ via local filtering operations $F_{optimal}$ and $E_{optimal}$ on their subsystems, respectively.  Thus, we can conclude the following lemma.

\begin{lemma}
	 The concurrence of the two-qubit state in Eq.~(\ref{eq:2qubitstate}) can be increased with ratio $\frac{1}{\sqrt{(1 - \| \vec{a} \|^2)(1 - \| \vec{b} \|^2)}}$  by applying optimal local filtering operations on each side, respectively.
\end{lemma}

\section{Effect of local measurement on concurrence}\label{Sec:effectofMeasurement}
In this section, we investigate the concurrence change under local measurement and try to answer the question of finding optimal quantum measurement to enhance quantum entanglement. We evaluate the performance of a measurement strategy via the expectation of the post-measurement concurrence which is denoted as $EC(\mathcal{M})$ for measurement apparatus $\mathcal{M}$. 

Suppose measurement  apparatus $\mathcal{M}\equiv \{M_k\}$ and  $\mathcal{N}\equiv \{N_j\}$ are performed on the subsystem of Alice and Bob, respectively. Trivial analysis shows that the expectation of post-measurement concurrence is as follows,
\begin{displaymath}
	\begin{array}{ccl}
		EC(\mathcal{M} \otimes \mathcal{N}) & = &  \sum_{k,j} p(M_k, N_j) C \frac{|det(M_k)|\cdot |det(N_j)|}{tr()(M_k^{\dagger}M_k \otimes N_j^{\dagger}N_j) \rho_{AB})}  \ \\
		& = & C \sum_k |det(M_k)| \cdot \sum_j |det(N_j)|  \\
		& = & \frac{1}{C} EC(\mathcal{M}) \cdot EC(\mathcal{N}),
	\end{array}
\end{displaymath}
which indicates that the  local measurements $\mathcal{M}\equiv \{M_k\}$ and  $\mathcal{N}\equiv \{N_j\}$ are independent with respect to the effect on entanglement. Thus, we only need to consider the performance on one side. 

We consider dichotomic measurement $\mathcal{M} \equiv \{M_0, M_1\}$ on Alice system. Suppose $M_{0}^{\dagger}M_{0} = \sum_{k=0}^1 m_{k}^{2} |m_k \rangle \langle m_k|$ is the spectral decomposition of the operator $M_{0}^{\dagger}M_{0}$. Due to the completeness of quantum measurement, we have $M_{1}^{\dagger}M_{1} = \sum_{k=0}^1 (1 - m_{k}^{2})  |m_k \rangle \langle m_k|$. Note that $|det(M_{0}) | = \sqrt{det(M_{0}^{\dagger}M_{0} )} = |m_0 m_1|$ and similarly $|det(M_{1})|  =  \sqrt{(1 - m_{0}^{2}) (1 - m_{1}^{2})}$. Thus, the expectation of concurrence is
\begin{displaymath}
	 \begin{array}{ccl}
	 	  EC(\mathcal{M}) & = & p(M_0) C(\rho_{AB}^{M_0}) +  p(M_1) C(\rho_{AB}^{M_1}) \\
	 	  		& = &C( |det(M_0)| +  |det(M_1)| )  \\
	 	  		& = & C( |m_0 m_1| + \sqrt{(1 - m_{0}^{2}) (1 - m_{1}^{2})}).
	 \end{array}
\end{displaymath}
Let $m_0 \equiv \sin{\theta}$ and $m_1 \equiv \sin{\phi}$ with $\theta, \phi \in[0, \frac{\pi}{2}]$. Then, it follows that
\begin{align}
	EC(\mathcal{M}) =  C(\sin{\theta} \sin{\phi} + \cos{\theta} \cos{\phi}) =C  \cos{(\theta - \phi)} .
\end{align}
Obviously,  $EC(\mathcal{M})  = 0$ when $\theta = \frac{\pi}{2} + \phi$, namely $\mathcal{M}$ is a projective measurement in the basis $\{ | m_0 \rangle, | m_1\rangle \}$, which coincides with the fact that projective measurement breaks entanglement in joint system.  And $EC(\mathcal{M})  = C$ when $\theta = \phi$, namely, the measurement operator is identity. It coincides with the fact that local operations can not create entanglement and therefore the best local operation for keeping quantum entanglement is the identity operation to leave it along.

Although the results are well known, our analytical  work gives more rigorous support to the facts.

\section{Conclusion}
In this article, we considered the problem of quantitatively increasing the  entanglement of a bipartite quantum state shared between two separated parties. Firstly, we investigated the optimal local filtering operation for increasing the most entanglement, measured by the concurrence of the joint system. With some non-trivial calculations, we derived the upper bound of the entanglement increase ratio and the analytical expression of the optimal operation for reaching the upper bound ratio as well as the success probability of the operation. We found that the optimal settings are completely determined by the corresponding local Bloch vector. Specifically, the entanglement increase ratio increases with the length of the Bloch vector while the success probability decreases with it. Secondly, we considered the effect of the general measurement by extending the result of the filtering operation. We found that any local measurement could not increase the expectation of entanglement. Although it is a well-known fact, our  result gives more rigorous analytical  evidence .

\begin{acknowledgments}
    This research was partially supported by the Innovation Program for Quantum Science and Technology (Grant No. 2021ZD0302900), National Natural Science Foundation of China (Grants No. 62002333), Fundamental Research Funds for the Central Universities (Grant No. WK2060000018), Anhui Initiative in Quantum Information Technologies (Grant No. AHY150100), and ANSO Scholarship for Young Talents.
\end{acknowledgments}


\bibliographystyle{unsrt}
\bibliography{references}

\end{document}